\documentclass[numberedappendix]{emulateapj}

\newcommand{\ds}{$\delta$~Scuti}
\newcommand{\da}{RRab}
\newcommand{\dc}{RRc}

\newcommand{\dq}{QSO}
\newcommand{\vkms}{km s$^{-1}$}
\newcommand{\kc}{$\cdots$}

\slugcomment{AJ accepted}

\begin{document}

\title{ The Identification of RR Lyrae and \ds~stars from variable $GALEX$ Ultraviolet Sources.}

\author{T.D. Kinman\altaffilmark{1}          }
\affil{National Optical Astronomy Observatory, 950 N. Cherry Avenue, Tucson, AZ 85719, USA}

\author{ Warren  R. Brown}                             
\affil{Smithsonian Astrophysical Observatory, 60 Garden St., Cambridge, MA 02138, USA}

\altaffiltext{1}{ 
NOAO is operated by AURA, Inc.\ under cooperative 
agreement with the National Science Foundation.}

\begin{abstract}

 We identify the RR Lyrae and \ds~stars in three catalogs of $GALEX$ variable 
 sources.  The $NUV$ amplitude of RR Lyrae stars is about twice that in $V$, so we 
 find a larger percentage of low amplitude variables than catalogs such as Abbas et 
 al. (2014).  Interestingly, the $(NUV - V)_{0}$ color is sensitive to metallicity 
 and can be used to distinguish between variables of the same period but differing 
 [Fe/H]. This color is also more sensitive to T$_{eff}$ than optical colors and can 
 be used to identify the red edge of the instability gap.
 We find 8 \ds~stars, 17 \dc~stars, 1 RRd star and 84 \da~stars in the 
 $GALEX$ variable catalogs of Welsh et al. (2005) and Wheatley et al. (2008).  We 
 also classify 6 \ds~stars, 5 \dc~stars and 18 \da~stars among the 55 variable 
 $GALEX$ sources identified as ``stars'' or RR Lyrae stars in the catalog of Gezari 
 et al. (2013).  We provide ephemerides and light curves for the 26 variables that 
 were not previously known.

\end{abstract}


\section{Introduction}

 The $Galaxy~Evolution~Explorer$ ($GALEX$) (Martin et al. 2005) has imaged the whole 
 sky in the $FUV$ ($\lambda\lambda$ 1350 -- 1750 \AA) and $NUV$ ($\lambda\lambda$ 
 1750 -- 2750 \AA) wavebands, opening the door to studying variable stars in the 
 ultraviolet.  One of the most important classes of variable stars are RR Lyraes, 
 evolved horizontal branch stars that have intrinsic luminosities that correlate 
 with pulsation amplitude and period.  \ds~stars (DSCT\footnote {These pulsating 
 variables (spectral types A0 to F5) have $V$ amplitudes in the range 0.03 to 0.90 
 mag.; those with $V$ amplitudes greater than 0.20 are often called High Amplitude
 \ds~stars.}) are a
 different class of pulsators with colors and pulsation periods comparable to RR 
 Lyraes; they are discussed in the appendix (Sec. A). Importantly, the ultraviolet 
 ($UV$) amplitudes of RR Lyrae stars are much larger in $GALEX$ wavebands than in 
 optical wavebands (Wheatley et al. 2003). What remains unexplored are complete 
 samples of $UV$ variables.

  Interestingly, detailed observations of several bright RR Lyrae stars show that 
 their $UV$--optical colors correlate with metallicity. Wheatley et al. (2012)
 fit model light-curves to the $FUV$ and $NUV$ photomtery of six well-observed 
 bright RR Lyrae stars. They find that the $FUV$ and $NUV$ light-curves, which 
 primarily reflect temperature changes during the pulsation cycle, also depend on 
 metallicity.  We also find that $(NUV - V)$ depends on metallicity. 

 To date, candidate variable $GALEX$ sources have been cataloged by Welsh (2005) 
 (Catalog 1) and by Wheatley et al. (2008) (Catalog 2).  During the course of this 
 work, a third catalog of over a thousand variable $GALEX$ sources was published by 
 Gezari et al. (2013); this catalog exclusively uses $NUV$ magnitudes.  Few of these 
 $UV$ variables have optical identifications.  Only 25\% of the 84 sources in 
 Catalog 1 and less than a third of the 410 sources in Catalog 2 are matched to 
 optical objects of known type.  Most of the $UV$ variables with optical 
 identifications are extragalactic objects.  Less than 10\% of the $UV$ variables in 
 the catalog of Gezari et al. (2013) have identifications with known types of stars.

 The goal of this paper is to identify all of the Galactic stars in the $GALEX$ 
 variable catalogs in a complete and systematic way, with a particular focus on RR 
 Lyrae variables.  We investigate the $(NUV - V)_{0}$ color index as a possible 
 diagnostic for identifying variable types:  in Sec. 4.5 and Appendix B, we address 
 the problem of distinguishing between type c RR Lyrae stars and contact binaries 
 (Kinman \& Brown 2008). We base our analysis on $NUV$ magnitudes, because $NUV$ 
 magnitudes are more accurate than $FUV$ magnitudes and are available for a larger 
 number of objects.  The use of the $(FUV - NUV)$ color could provide additional  
 constraint (see for example, Smith et al. 2014) but would limit the sample.

\begin{figure}
\plotone{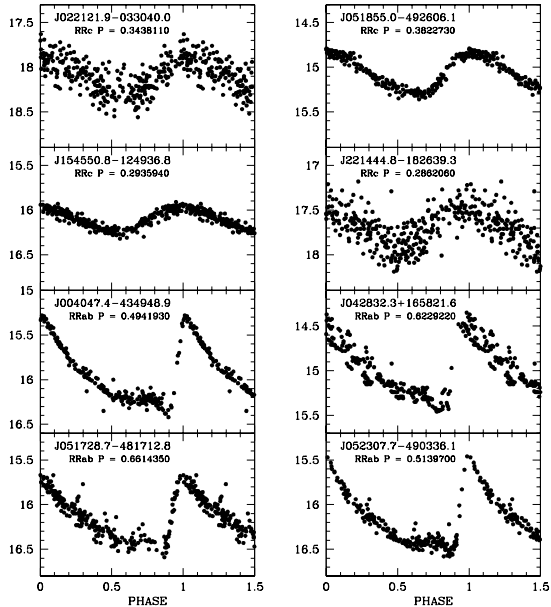}
\caption {$V$-magnitude light curves for 4 newly identified RRc and 4 newly 
	identified RRab variables presented in Table 2. 
                     }
\label{Fig1}
\end{figure}

\begin{figure}
\plotone{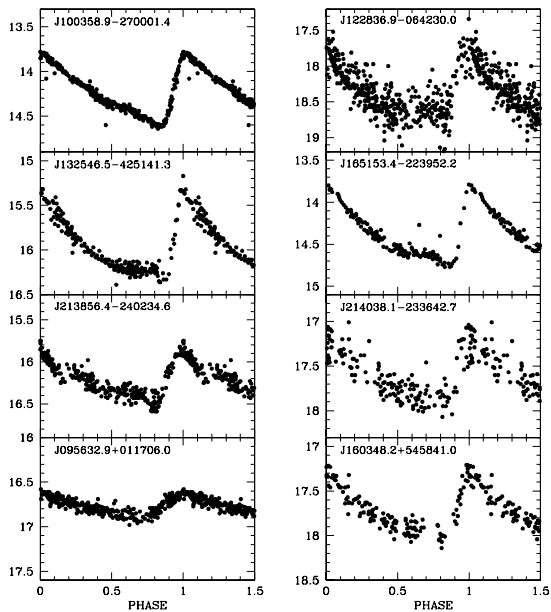}
\caption { $V$-magnitude light curves for 8 new identified RRab variables 
	presented in Table 2.
                     }
\label{Fig2}
\end{figure}

\begin{figure}
\plotone{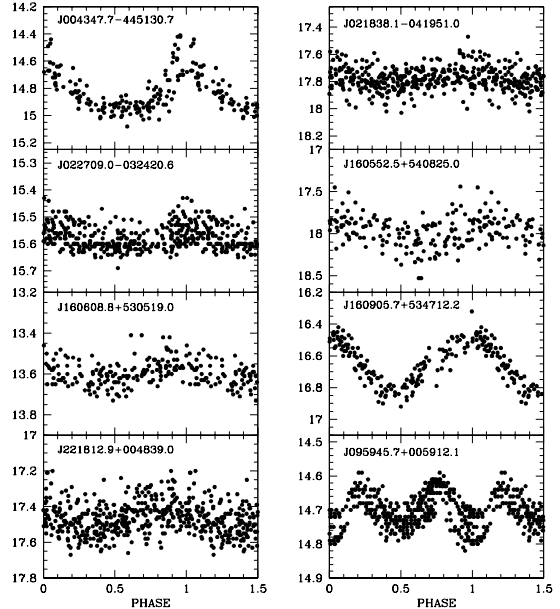}
\caption { $V$-magnitude light curves for 7 newly identified \ds~stars and one EW 
      variable presented in Table 2.  The periods of the lowest amplitude \ds~stars 
      are quite uncertain.
                     }
\label{Fig3}
\end{figure}

\begin{figure}
\plotone{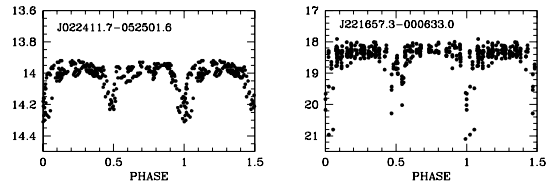}
\caption { $V$-magnitude light curves for 2 newly identified Algol (EA) variables 
    presented in Table 2. The period of J221657.3$-$000633.0 is uncertain.
                     }
\label{Fig4}
\end{figure}

 Our approach is to use the VizieR search tool and the JAAVS variable star catalog 
 (VSX, Watson 2006) to identify known variables, such as those identified by Drake 
 et al. (2012) in the Catalina Surveys, by Palaversa et al. (2013) in the LINEAR 
 Survey, and by Abbas et al. (2014) in their search of the combined SDSS, Pan-STARRS1
 and Catalina Surveys\footnote{The Abbas et al. Catalog was published only shortly 
 before this paper was completed and so it has only been possible to make limited use
 of it.}.
 We look at every object in Catalog 1 and in Catalog 2, and the subset of 
 objects identified as ``stars'' in the larger Gezari catalog.  If we are unable to 
 classify an object with the variable star catalogs, we then use data from the 
 Catalina Survey Data Release 2 (Drake et al. 2009) to measure the object's mean $V$ 
 magnitude and, if possible, its variability.  Period-finding with the Catalina 
 Survey data (e.g. using the NASA Exoplanet Archive Periodogram Service) becomes 
 difficult for objects with $V$ $>$ 17 and for objects with $V$ amplitudes 
 $\lesssim$0.2 magnitudes; consequently, the types we assign to such variables are 
 less certain.  We also review the 38 objects previously identified as RR Lyrae 
 stars in the Gezari catalog.

 This paper presents the RR Lyrae and \ds~stars that we identify, and describes how 
 $GALEX$ magnitudes can assist in their identification.  The $(NUV-V)_{0}$ color is 
 particularly sensitive for detecting the temperature changes that come from 
 pulsations, and for detecting the onset of this instability as a function of 
 parameters such as T$_{eff}$.  The identification of other types of stars and 
 variables from the $GALEX$ variability catalogs will be discussed elsewhere.

\section {The identification of RR Lyrae and \ds~Stars in the first two $GALEX$ variability Catalogs.
}
    Table 1 presents 110 RR Lyrae and \ds~stars that we identify from the first two 
    catalogs of $GALEX$ variables; these stars constitute 22\% of all the sources in 
    these catalogs. Our list comprises 8 \ds~stars, 17 \dc~stars, 1 RRd star and 84 
    \da~stars.  References to the sources of the $GALEX$ magnitudes and the optical 
    data are given in columns 4 and 10 respectively. The $(NUV - V)_{0}$ colors 
    given in col. 8 refer to mean magnitudes \footnote {The mean $NUV$ magnitude is the mean of the maximum and minimum values. The mean $V$ magnitude is the arithmetic    mean of all the observations given by the Catalina DataRelease 2. For brighter   variables not in this survey, the mean was estimated from the maximum and minimum magnitudes using the formula given in the appendix of Preston et al. (1991).}  and their largest source of uncertainty 
    lies in the estimation of the mean $NUV$ from the available data; this 
    uncertainty increases with the $NUV$ amplitude and is therefore largest for the 
    RRab variables. We assumed the galactic extinction A$(NUV)$ = 8.90E$(B-V)$ (Rey 
    et al. 2007). The E$(B-V)$ were taken from the MAST archive of $GALEX$ objects. 
    Only 6 stars have E$(B-V)$ $>$ 0.100 so the uncertainty in the correction for 
    extinction (Peek \& Schiminovich, 2013) is probably significant only for these 
    stars. The periods given in column 9 are those taken from the literature in 
    the case of previously known variables; otherwise, the periods were 
    determined by us. 

    We discover 2 new \ds~stars, 3 \dc~stars and 10 \da~stars among the 110 variable 
    stars.  We present the ephemerides of the new variables in Table 2, and plot 
    their light curves in Figures 1, 2, 3 \& 4. One of these \dc~and one of 
    these \da~ were later found in the catalog of Drake et al, (2014) but
    with differing periods; these differences are discussed in Sec. 4.6.

\begin{figure}
\plotone{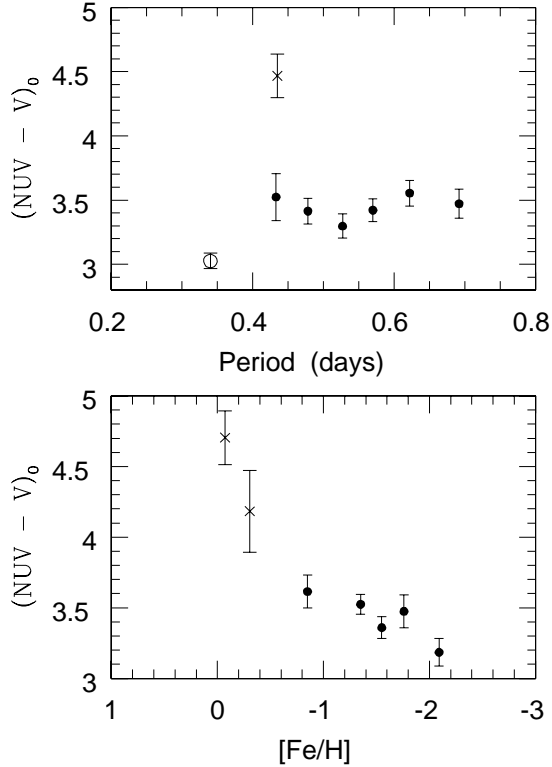}
\caption { 
  Distribution of known RR Lyrae stars from Dambis et al. (2013) in plots of the de-reddened 
  $(NUV - V)_{0}$ color versus period (upper panel) and versus [Fe/H] (lower panel).
  The filled circles, crosses, and open circles denote halo RRab, disk RRab, and
  RRc variables, respectively.
  There are significant differences in $(NUV - V)_{0}$ between the disk RRab stars 
  ([Fe/H] $\geq -0.40$), the halo RRab stars ([Fe/H] $\leq -0.80$), and the RRc
  stars.
                     }
\label{Fig5}
\end{figure}

    We investigate trends between $(NUV-V)_{0}$ color and stellar properties by 
    studying a well-defined sample of known RR Lyrae stars.  We begin by calculating 
    $(NUV-V)_{0}$ for 150 nearby RR Lyrae stars using the recent compilation by 
    Dambis et al. (2013), a sample for which we could easily obtain the necessary 
    data.  For this sample, we plot $(NUV-V)_{0}$ against period for RRc and for 
    $halo$ and $disk$ RRab (Fig 5, upper panel), and against [Fe/H] (Fig 5, lower 
    panel).

    There is a significant difference in $(NUV-V)_{0}$ between the RRc and 
    RRab that reflects the average temperature difference between the two types. 
    There is also a significant difference in $(NUV-V)_{0}$ between the $disk$ and 
    $halo$ RRab that reflects the metallicity difference between the two types as 
    expected from Wheatley et al. (2012).

    We also investigate trends between the $(NUV-V)_{0}$ color and stellar properties 
    for a sample of known high-amplitude \ds~stars (HADS) taken from McNamara (1997) 
    and \ds~ of lower amplitude taken from Suveges et al. (2012). The results are 
    shown in the plot of $(NUV-V)_{0}$ $vs.$ Period given in Fig. 6. The blue open 
    circles, black open triangles and black filled triangles are respectively 
    the RRc (both Catalogs), RRab from Catalog 1 and RRab from Catalog 2 in 
    Table 1. The horizontal solid blue, dotted black and dashed red lines show the mean values of 
    $(NUV-V)_{0}$ for the RRc, halo RRab, and disk RRab respectively; their 
    associated vertical lines show the $r.m.s.$ standard errors of these means. 
    There is good agreement between these mean values and the $(NUV-V)_{0}$ of our 
    program stars.  The observed increase in this color with increasing period is 
    nicely matched to the $(NUV-V)_{0}$ location of the blue edge of the instability 
    gap (Kinman et al. 2007), shown by the heavy blue vertical line at a period of 
    0.2 days in Fig. 6.

\begin{figure}
\plotone{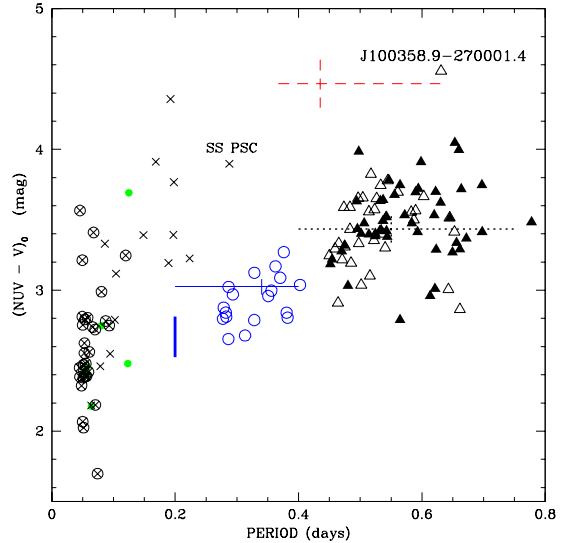}
\caption {
  De-reddened $(NUV - V)_{0}$ color versus pulsation period for our program stars
  from Table 1. The stars are plotted as:  high-amplitude \ds~stars (green
  solid circles), RRc (blue open circles), RRab from Catalog 1 (black open triangles)
  and RRab from Catalogue 2 (black filled triangles). The blue (solid), black
  (dotted)  and red  (dashed) horizontal lines show mean values of $(NUV 
  - V)_{0}$ for known RRc, halo RRab, and disk RRab variables, respectively.  Two 
  samples of known high-amplitude \ds~stars variables are plotted with black x's 
  (McNamara 1997) and encircled black x's (Suveges et al. 2012). The heavy blue 
  vertical line shows the range of $(NUV - V)_0$ for blue horizontal branc stars at 
  the blue edge of the instability gap (Kinman et al. 2007).
                     }
\label{Fig6}
\end{figure}

\begin{deluxetable*}{lccccccclll}
\tablewidth{0cm}
\tabletypesize{\scriptsize}
\tablecaption{Identifications of \ds ~(DCST) and RR Lyrae Stars in Catalogs 1 and 2.
 [Available in full in the Online Journal]}
\tablehead{ 
\colhead{ Star \tablenotemark{a}     } &
\colhead{$NUV_{max}$ \tablenotemark{b}     } &
\colhead{$NUV_{amp}$ \tablenotemark{c}     } &
\colhead{ Ref \tablenotemark{d}    }  &
\colhead{E$(B-V)$      } &
\colhead{$V_{max}$ \tablenotemark{e}     } &
\colhead{$V_{amp}$ \tablenotemark{f}     } &
\colhead{$(NUV-V)_{0}$ \tablenotemark{g}     } &
\colhead{ Period \tablenotemark{h}     } &
\colhead{ Refs \tablenotemark{i}    }  &
\colhead{ Notes    } 
}
\startdata
 DCST:             &      &    & &     &     &    &     &         &     &     \\
 J004347.9+421654.9& 13.02&0.63&1,2&0.084&09.19&0.27&3.692&0.1249078&13   & {\bf 1} \\ 
 J004347.7--445130.7&16.78&0.84&2&0.009&14.45&0.50&2.457&0.0574790&20   & {\bf 2} \\
 J033715.2--283238.1&15.51&0.94&2&0.011&13.20&0.65&2.397&0.0561380&11   &         \\
 J104647.6+591445.5& 19.86&0.93&2&0.008&17.40&0.80&2.480&0.1231520&20   & {\bf 3} \\ 
 J133757.1+401610.8& 20.00&0.91&1&0.006&18.00&0.50&2.177&0.0641760&16   &         \\
 J160608.8+530519.0& 18.76&1.53&2&0.014&13.53&0.55&5.647&0.1027590&20   & {\bf 2} \\
 J164106.8+404225.2& 15.70&0.87&2&0.008&13.40&0.55&2.419&0.0512960&13   & {\bf 4} \\
 J171250.9+582748.6& 18.51&0.63&2&0.044&15.70&0.30&2.749&0.0803930&16,7 & {\bf 5} \\ 
 RRc:             &      &    &       &     &    &     &         &      &         \\
 J033546.0--290209.8&16.09&0.90&2&0.008&13.45&0.50&2.797&0.2773379&18   &         \\
 J051855.0--492606.1&17.59&0.84&2&0.026&14.85&0.45&2.806&0.3822732&20   & {\bf 2} \\ 
 J103720.9+591653.8& 18.31&0.62&2&0.008&15.70&0.40&2.679&0.3132750&10,21,22&         \\
 J103759.4+581213.5& 19.67&1.03&2&0.008&16.90&0.40&3.038&0.4023800&21   & {\bf 2} \\ 
 J105044.4+045115.9& 19.51&0.97&2&0.027&16.80&0.45&2.813&0.2827820&16,19.22&         \\
 J143823.9+635329.9& 19.23&0.76&2&0.016&16.34&0.36&2.998&0.3559790&10,21&         \\
 J144033.3--781914.6&17.09&0.79&2&0.108&13.80&0.70&2.789&0.3281820&13   & {\bf 6} \\ 
 J154550.8--124936.8&19.51&0.62&2&0.146&15.96&0.28&2.970&0.2935940&20,22 & {\bf 2, 45} \\
 J155800.6+535233.6& 18.06&0.87&2&0.012&15.12&0.36&3.124&0.3284030&16,10,21,22&        \\
 J160759.9+535209.9& 19.20&0.90&2&0.012&16.52&0.44&2.841&0.3806390&12,22   &         \\
 J171008.9+585112.9& 20.06&0.92&2&0.031&17.25&0.50&2.842&0.2821010&16,21,22&         \\
 J203922.4--010346.0&16.88&0.81&2&0.084&13.57&0.40&3.089&0.3703490&14,7,22 & {\bf 7} \\
 J204024.3--010950.0&19.06&0.72&2&0.079&15.68&0.34&3.170&0.3626603&14,7,21& {\bf 8} \\
 J204042.8--003717.3&20.28&0.72&2&0.159&17.12&0.40&3.023&0.2865740&14,21&     \\
 J221444.8--182639.3&20.00&0.96&2&0.026&17.40&0.55&2.654&0.2862060&20   & {\bf 2} \\
 J223601.0+134711.5& 19.16&0.91&2&0.068&15.80&0.40&3.271&0.3760710&16,7,22 & {\bf 9}\\
 J224029.0+120056.3& 17.56&0.85&2&0.050&14.60&0.50&2.877&0.2786230&16,7,21,22& {\bf 10}\\
 RRd:             &      &    &       &     &    &     &         &      &         \\
 J130204.5+463533.7& 18.01&1.45&1&0.014&15.45&0.60&2.959&0.3509191&12   & {\bf 11}\\ 
 RRab:             &      &    &       &     &    &     &         &     &         \\
 J001611.6--392721.5&16.16&2.02&2&0.015&13.50&1.00&3.420&0.5316009&11   &         \\
 J003043.0--422747.7&17.25&1.48&2&0.007&14.00&0.90&3.747&0.6973440&9    &         \\
 J004047.4--434948.9&18.20&1.95&2&0.016&15.30&0.95&3.633&0.4941930&20   & {\bf 2} \\
 J004548.2--435509.1&16.00&1.38&1&0.001&13.00&0.70&3.565&0.5898310&11   &         \\
 J011010.9--452431.3&18.84&1.85&2&0.009&15.90&0.85&3.697&0.5897626&13   & {\bf 12}\\
 J011805.2--751126.1&18.61&1.41&2&0.055&15.60&1.20&3.032&0.480    &13   & {\bf 13}\\
 J013642.0--062743.1&19.70&0.63&2&0.028&16.35&0.25&3.483&0.7775660&16,7,22 & {\bf 14}\\ 
 J022521.8--050017.4&19.42&1.57&2&0.027&16.90&0.80&3.012&0.6212419&3,21 &         \\ 
 J033500.4--272854.8&16.98&1.63&2&0.009&14.20&0.80&3.415&0.6975764&18   &         \\
 J042832.3+165821.6& 19.30&1.64&2&0.377&14.42&1.00&3.288&0.6229220&20,7 &{\bf 2,15}\\
 J051728.7--481712.8&18.60&1.42&2&0.027&15.70&0.80&3.291&0.6614350&20   &{\bf 2}  \\
 J052307.7--490336.1&18.17&2.13&2&0.030&15.47&1.10&3.397&0.5139700&20   &{\bf 2}  \\
 J081226.4+033320.5& 18.95&1.95&2&0.023&16.08&0.72&3.677&0.5555151&6,7  &{\bf 16} \\
 J093026.0+071221.6& 15.83&0.97&1&0.058&12.08&0.79&3.666&0.6028494&13   &{\bf 17} \\
 J100133.2+014328.5& 17.21&1.98&2&0.019&14.40&1.00&3.521&0.5423741&3,10,19,21&    \\
 J100358.9--270001.4&18.26&1.43&1&0.079&13.78&0.84&4.556&0.6311440&20   &{\bf 2}  \\ 
 J102002.7+611538.9& 16.71&1.62&1&0.007&14.34&0.81&3.005&0.6428627&5,3,10,21&        \\
 J102641.8+572858.7& 18.78&1.84&2&0.011&15.80&0.85&3.719&0.6636605&10,21&         \\
 J103538.2+581549.1& 15.53&1.92&2&0.008&12.87&1.02&3.384&0.5242549&13,7 &{\bf 18}\\ 
 J104707.6+053349.2& 17.78&2.25&2&0.024&15.00&1.00&3.642&0.5369295&10,19,21&      \\
 J104803.1+554209.9& 18.32&2.14&2&0.008&15.70&1.05&3.476&0.5065280&10,21&         \\
 J104844.2+581538.6& 18.47&2.49&2&0.011&16.10&1.30&3.317&0.4748198&3,21 &         \\
 J105130.9+043801.7& 17.74&2.03&2&0.033&15.00&1.00&3.403&0.5243170&16,10,19,21&    \\
 J105513.8+564747.7& 17.09&2.24&2&0.007&14.65&0.75&3.519&0.5417451&16,10&         \\
 J105622.3+570522.1& 17.09&2.27&2&0.009&14.70&0.90&3.402&0.5016290&3,10,21&         \\
 J105926.1--005927.9&19.53&2.09&1&0.042&16.72&1.43&3.246&0.4499016&13,3,42&{\bf 19}\\
 J111147.3+510549.4& 18.12&1.50&1&0.014&15.10&0.85&3.514&0.6449104&3    &          \\
 J112334.9+474014.6& 19.45&1.07&1&0.014&16.18&1.10&3.353&0.5224715&3,10,21&         \\
 J113340.3+502328.0& 16.71&2.93&1&0.014&14.70&1.10&3.333&0.4976910&13,21& {\bf 20}\\
 {\bf Truncated for display} &      &    &       &     &    &     &   &   &
\enddata

\tablenotetext{a}{$GALEX$ identification of source.}           
\tablenotetext{b}{$NUV$ magnitude at maximum.}
\tablenotetext{c}{Amplitude of $NUV$ magnitude. }
\tablenotetext{d}{Reference for $GALEX$ source.}
\tablenotetext{e}{$V$ magnitude at maximum.}
\tablenotetext{f}{Amplitude of $V$ magnitude.}
\tablenotetext{g}{Difference beteween mean $NUV$ and $V$ magnitudes corrected for galactic extinction.}
\tablenotetext{h}{Period in days.}
\tablenotetext{i}{References for optical data.}

\tablerefs{
{(1)}~Welsh+(2005)(The $GALEX$ Ultraviolet Variability Catalog);~
{(2)}~Wheatley+(2008)(The Second $GALEX$ Ultraviolet Varaiability Catalog);~
{(3)}~Drake+(2013);~                      
{(4)}~Keller+(2008);~                       
{(5)}~Kinemuchi+(2006);~                    
{(6)}~Kraus+(2007);~                    
{(7)}~Li+(2011);~                    
{(8)}~Miceli+(2008);~                    
{(9)}~Norton+(2007);~                    
{(10)}~Palaversa+(2013);~                    
{(11)}~Pojmanski(2002)(ASAS Catalogue);~                    
{(12)}~Poleski(2013);~                    
{(13)}~Samus+(2012)(Gen. Cat. Variable Stars);~                    
{(14)}~Sesar+(2010);~                    
{(15)}~Vivas+(2004)(QUEST Survey);~                    
{(16)}~Watson+(2006)(AAVSO VSX Catalog);~                    
{(17)}~Wils+(2006);~                    
{(18)}~Wils(2010)(Data in VSX Catalog);~                    
{(19)}~Zinn+(2014)(Extension of QUEST Survey);~                    
{(20)}~This Paper~(New Identification). 
{(21)}~Abbas+ (2014);                 
{(22)}~Drake+ (2014).
}


\tablecomments{
{\bf (1)}~CC And.~                                
{\bf (2)}~New identification. Ephemeris in Table 2.~
{\bf (3)}~New Identification. Type and period uncertain.~
{\bf (4)}~v1209 Her (SX Phe type).~                             
{\bf (5)}~Sp~Type A (7)~                             
{\bf (6)}~v344 Aps.~                             
{\bf (7)}~Sp~Type A (7)~                             
{\bf (8)}~Sp~Type F (7)~                             
{\bf (9)}~Sp~Type A (7)~                             
{\bf (10)}~Sp~Typ A (7)~                             
{\bf (11)}~Period is first overtone.~                             
{\bf (12)}~UU Phe.~                                
{\bf (13)}~BG Tuc, B magnitude.~                               
{\bf (14)}~Sp~Typ A (7)~                             
{\bf (15)}~Sp~Typ F (7)~                             
{\bf (16)}~Sp~Typ F (7)~                             
{\bf (17)}~WW Leo.~                               
{\bf (18)}~v341 Uma;Sp~Type A (7).~                             
{\bf (19)}~IX Leo.~                               
{\bf (20)}~CZ Uma.~                               
}

\end{deluxetable*}

\begin{deluxetable*}{lccccccc}
   \tabletypesize{\scriptsize}
   \tablewidth{0pt}
   \tablecaption{ Positions, Types and Ephemerides of New$^{\dagger}$  Variables. }
   \tablehead{  
\colhead{ Star \tablenotemark{a}     } &
\colhead{ R.A. \tablenotemark{b}     } &
\colhead{ Dec. \tablenotemark{c}     } &
\colhead{ Type \tablenotemark{d}     } &
\colhead{ HJD(max) \tablenotemark{e}     } &
\colhead{ Period \tablenotemark{f}     } &
\colhead{ $V_{max}$ \tablenotemark{g}     } &
\colhead{ $V_{min}$ \tablenotemark{h}     }\\ 
                     & (2000) & (2000)   &    &2450000.0+&(days) &(mag.)&(mag.) \\
 }
  \startdata
 CATALOGS 1 and 2:   &        &          &    &         &        &     &       \\
 J004347.7$-$445130.7&010.9488&$-$44.8586&DCST&3556.1599&0.057479&14.45&14.95  \\
 J160608.8$+$530519.0&241.5371&$+$53.0890&DCST&3656.5589&0.102759&13.53&13.64  \\
                     &        &          &    &         &        &     &       \\
 J051855.0$-$492606.1&079.7293&$-$49.4349& RRc&3578.5038&0.382273&14.85&15.30  \\
 J154550.8$-$124936.8$^{\ddagger}$&236.4618&$-$12.8269& RRc&3523.4753&0.293594&15.96&16.24  \\
 J221444.8$-$182639.3&333.6870&$-$18.4444& RRc&3583.6795&0.286206&17.40&17.95  \\
                     &        &          &    &         &        &     &       \\
 J004047.4$-$434948.9&010.1981&$-$43.8301&RRab&3555.8138&0.494193&15.30&16.25  \\
 J042832.3$+$165821.6&067.1349&$+$16.9726&RRab&3469.5764&0.622922&14.45&15.45  \\
 J051728.7$-$481712.8&079.3700&$-$48.2869&RRab&3576.4696&0.661435&15.70&16.50  \\
 J052307.7$-$490336.1&080.7823&$-$49.0599&RRab&3576.4712&0.513970&15.47&16.57  \\
 J100358.9$-$270001.4&150,9955&$-$27.0002&RRab&3694.8098&0.631144&13.78&14.62  \\  
 J122836.9$-$064230.0$^{\ddagger}$&187.1540&$-$06.7085&RRab&3496.5066&0.501648&17.50&18.60  \\  
 J132546.5$-$425141.3&201.4437&$-$42.8615&RRab&3556.7107&0.451759&15.25&16.25  \\
 J165153.4$-$223952.2&252.9728&$-$22.6644&RRab&3599.5306&0.613432&13.80&14.75  \\
 J213856.4$-$240234.6&324.7351&$-$24.0430&RRab&3597.8294&0.594016&15.80&16.55  \\
 J214038.1$-$233642.7&325.1592&$-$23.6118&RRab&3638.7051&0.564344&17.25&17.90  \\
 Gezari et al. (2013):&        &          &    &         &        &     &       \\
 J022709.0$-$032420.6&036.7872&$-$03.4058&DCST&3648.7343&0.1685980&16.60&16.69 \\
 J221812.8$+$004839.0&334.5535&$+$00.8106&DCST&3507.9368&0.0413220&17.35&17.50 \\
 J160552.5$+$540825.0&241.4690&$+$54.1404&DCST&3880.8828&0.0587400&17.70&18.10 \\
 J160905.7$+$534712.2$^{\ddagger}$&242.2740&$+$53.7866&DCST&3505.5268&0.1697210&16.50&16.85 \\
 J021838.1$-$041951.0&034.6589&$-$04.3304&DCST&3627.8700&0.0759360&17.70&17.80 \\
                     &        &          &    &         &        &     &       \\
 J022121.9$-$033040.0$^{\ddagger}$&035.3413&$-$03.5109&RRc &3627.5444&0.3438110&17.90&18.30 \\
                     &        &          &    &         &        &     &       \\
 J095632.9$+$011706.0&149.1372&$+$01.2850&RRab&3464.1026&0.7711330&16.60&16.69 \\
 J160348.2$+$545841.0&240.9508&$+$54.9780&RRab&3880.6148&0.5418940&17.25&18.05 \\
                     &        &          &    &         &        &     &       \\
 J095945.7$+$005912.1&149.9400&$+$00.9866&EW  &3464.4354&0.2874170&14.62&14.75 \\
                     &        &          &    &         &        &     &       \\
 J221657.3$-$000633.0&334.2392&$-$00.1091&EA  &3502.9800&18.32325&18.30&20.:   \\
 J022411.7$-$052501.6$^{\ddagger}$&036.0485&$-$05.4172&EA  &3627.8052&0.8254050&13.95&14.3: 
   \enddata

\vspace{1mm}

\tablenotetext{$\dagger$}{All stars in this table had not been identified as 
variables prior to 2014 May 29.}

\tablenotetext{a}{$GALEX$ identification of source.}           
\tablenotetext{b}{Right Ascension in decimal degrees.}
\tablenotetext{c}{Declination in decimal degrees.} 
\tablenotetext{d}{Variable Type.}
\tablenotetext{e}{Heliocentric Julian Date of Maximum.}
\tablenotetext{f}{Period in days.}
\tablenotetext{g}{$V$ magnitude at maximum.}
\tablenotetext{h}{$V$ magnitude at minimum.}

\tablenotetext{$\ddagger$}{These stars were identified as variables by Drake et al. 
(2014).}

\end{deluxetable*}

   \begin{deluxetable*}{llcccccccccl}
   \tabletypesize{\scriptsize}
   \tablewidth{0pt}
   \tablecaption{ Data for the Variable Stars identified by Gezari et al. (2013) .}
   \tablehead{  
\colhead{ R.A. \tablenotemark{a} } &
\colhead{ Dec. \tablenotemark{b} } &
\colhead{ $NUV_{amp}$ \tablenotemark{c}     } &
\colhead{ LC\tablenotemark{d}     } &
\colhead{ $(NUV - V)_{0}$ \tablenotemark{e}     } &
\colhead{ Period \tablenotemark{f}     } &
\colhead{ $V$\tablenotemark{g}     } &
\colhead{ $(B-V)$ \tablenotemark{h}  } &  
\colhead{Type \tablenotemark{i}  }  &
\colhead{Type \tablenotemark{j}  }  &
\colhead{Type \tablenotemark{k}  }  &
\colhead{Notes  }  \\
  (2000) & (2000)&(mag.)&   & (mag.)& (days)&(mag.)&(mag.)&  &  &  &    \\
 }
  \startdata
  212.2576& +53.1997& 0.24&V& 3.298& $\cdots$ & 16.39& +0.52&RR&RR&\kc&     \\
   36.7721&--03.7063& 0.36&V& 1.896& $\cdots$ & 16.54& +0.26&RR&RR&\kc&     \\
   36.7872&--03.4058& 0.37&V& 2.103& 0.1685980& 15.51& +0.32&RR&RR&\ds&(1)(18)\\
   36.4632&--04.8917& 0.43&V& 1.912& $\cdots$ & 16.73& +0.19&RR&RR&\kc&     \\ 
  334.5535& +00.8106& 0.45&V& 1.728& 0.0413220& 17.30& +0.21&RR&RR&\ds& (1) \\ 
  150.0513& +01.0615& 0.46&V& 3,361& $\cdots$ & 17.19& +0.49&RR&RR&\kc&     \\ 
  150.0909& +01.2796& 0.49&V& 4.477& 0.3162890& 14.82& +0.61&RR&RR&EW & (2)(22)\\
   34.6594&--04.3309& 0.51&V& 2.280& 0.0759360& 17.72& +0.26&RR&RR&\ds& (1)  \\
  332.9349&--00.4230& 0.63&V& 1.782& $\cdots$ & 17.42& +0.50&RR&RR&\kc&     \\
  149.1372& +01.2850& 0.73&V& 3.359& 0.7711330& 16.69& +0.37&RR&RR&\da& (1) \\
  149.6863& +02.5218& 0.79&V& 3.553& 0.6468498& 17.61& +0.47&RR&RR&\da& (3) \\
  241.4690& +54.1404& 0.80& & 2.376& 0.0587400& 17.96& +0.32&RR&RR&\ds& (1) \\
  242.2740& +53.7866& 0.89& & 3.882& 0.1697210& 16.61& +0.54&RR&RR&\ds& (1)(23)\\
  241.9999& +53.8693& 0.95&V& 2.619& 0.3806406& 16.71& +0.20&RR&RR&RRab& (4,5)\\
   35.3413&--03.5109& 0.95&V& 2.820& 0.3438094& 18.04& +0.29&RR&RR&\dc& (1,7,22)\\
  333.3199&--00.5376& 0.95& & 2.114& 0.3365870& 16.66&--0.45&RR&RR&\dc& (6,21,22)  \\
  333.5874& +01.3271& 0.96& & 4.347& $\cdots$ & 17.22& +0.55&RR&RR&\kc&     \\
  332.4833&--00.6473& 0.99&V& 2.290& 0.3016423& 16.05& +0.22&RR&RR&\dc& (8,9,21,22) \\
  215.9569& +52.6509& 1.00&V& 0.961& $\cdots$ & 20.30& +0.53&RR&RR&\kc&     \\
  150.2291& +01.3947& 1.00&V& 2.651& 0.3210413& 15.75& +0.36&RR&RR&\dc& (2,21,22) \\
 216.3439&+52.7295&$>$1.04&F&$>$3.263&$\cdots$& 19.42&+0.46&RR&RR&\kc &     \\
  214.0191& +52.4641& 1.11&V& 3.532& 0.5825432& 16.41& +0.51&RR&RR&\da& (2,3)\\
   333.8419& +00.5817& 1.18&V&0.960& $\cdots$ & 20.07& +0.26&RR&RR&\ds~?& (10)\\
   212.8810& +53.5101& 1.29& & 3.375& 0.5807728& 16.54&+0.41&RR&RR&\da& (11,21,22) \\
   150.5272& +01.3757& 1.41&V& 3.271& 0.7289104& 16.28&+0.50&RR&RR&\da& (2,3,21)\\
   332.8058&--00.3874& 1.55&V& 2.612& 0.5807727& 16.98&+0.33&RR&RR&\da& (2,21) \\
  242.8745& +53.7296& 1.57&V& 3.052& 0.5838609& 16.07&+0.52&RR&RR&\da&(2,4,12,21)\\
   334.4712&--00.0933& 1.71&V& 2.876& 0.6556452& 13.93&+0.37&RR&RR&\da&(13) \\
   241.8077& +55.8589& 1.85& & 3.286& 0.5464062& 15.92&+0.18&RR&RR&\da&(2,3,4,21)\\
  241.2133& +54.1410& 1.89& & 2.997& 0.5161044& 16.55&+0.13&RR&BHB&\da&(2,3)\\
 334.2392&--0.1091&$>$1.91&F&$>$2.754&18.32325& 18.20&+0.24&RR&RR&EA &(1,14) \\
   150.3887& +01.7245& 1.93&V& 3.047& 0.5423790& 14.91&+0.26&RR&RR&\da&(2,3,4,21)\\
    36.3486&--05.5392& 2.07&V& 2.832& 0.5363961& 16.79&+0.46&RR&RR&\da&(3,20,21)\\
   240.9508& +54.9780& 2.11&V& 3.465& 0.5418940& 17.71&+0.24&RR&RR&\da&(1) \\
    36.3409&--05.0050& 2.30&V& 2.888& 0.6212305& 17.27&+0.46&RR&RR&\da& (3,21) \\
 150.1707&+01.3682&$>$2.70&V&$>$3.115& 0.4888790& 18.55&+0.16&RR&BHB&\da& (3) \\
   332.9690&   0.6330& 2.91&V& 2.633& 0.5053500& 16.73&+0.31&RR&RR&\da& (3,21) \\
 333.0054& --0.0571& 2.92&V& 2.680& 0.5030960& 17.42&+0.32&\dq&\da&\da& (6) \\
         &         &     & &      &          &      &     & &   &   &      \\
  36.0485&--05.4172& 1.03&F& 5.38 & 0.8254050& 13.94&0.89 &\kc&Star&EA &(1,15,24)\\
 150.8023& +00.9722& 1.00&F& 3.74 & 0.6061280& 16.08&0.55 &\kc&Star&\da&(3)(19)  \\
 333.0985& +00.0828& 0.96&V& 2.18 & 0.3505063& 15.85&0.17 &\kc&Star&\dc&(6,22)\\
 149.9732& +03.2086& 0.72&V& 4.91 & 0.2738640& 15.12&0.69 &\kc&Star&EA &(2,22)   \\
 215.2242& +54.4815& 0.61& & 5.24 & 0.3586440& 13.92&0.58 &\kc&Star&EW &(16,22) \\
  35.5465&--03.1702& 0.57&V& 4.63 & 0.4051308& 13.14&0.60&\kc&Star&EW &(17.15,22)\\
 149.9400& +00.9866& 0.27&V& 4.44 & 0.2874170& 14.63&0.66 &\kc&Star&EW &(1) 
   \enddata
\tablenotetext{a}{~Right Ascension in decimal degrees.}
\tablenotetext{b}{~Declination in decimal degrees.} 
\tablenotetext{c}{~Amplitude of $NUV$.}
\tablenotetext{d}{~Light curve type: V = Stochastic; F = Flaring (Gezari et al., 2013).}
\tablenotetext{e}{~Difference between mean $NUV$ and $V$ magnitudes corrected for galactic extinction.}
\tablenotetext{f}{~Period in days.}
\tablenotetext{g}{~Mean $V$ corrected for extinction.}
\tablenotetext{h}{~Mean $(B-V)$ corrected for extinction.}
\tablenotetext{i}{~Variable type from color (Gezari et al., 2013).}
\tablenotetext{j}{~Variable type (Gezari et al., 2013); RR = RR Lyrae, BHB = blue HB star.}
\tablenotetext{k}{~Variable type (This paper). EA are Algol-type eclipsing systems. EW are W Ursa Majoris-type eclipsing systems.}
 
\tablecomments{
{\bf (1)}~New Identification (this paper).~                                
{\bf (2)}~Palaversa et al. (2013).~
{\bf (3)}~Drake et al. (2012).~ 
{\bf (4)}~Wheatley et al. (2008) Second $GALEX$ catalogue of variables.~ 
{\bf (5)}~Drake et al. (2014) give type and period. Poleski (2014) gives same period.~                             
{\bf (6)}~Sesar et al. (2010).~                             
{\bf (7)}~Radial velocity --154$\pm$2 \vkms~(SDSS DR 9).~                     
{\bf (8)}~Sesar et al. (2012).~                             
{\bf (9)}~Radial velocity --213$\pm$1 \vkms~(SDSS DR 9).~                     
{\bf (10)}~New Identification. Possible HADS but period uncertain. ~  
{\bf (11)}~CL Boo. ~
{\bf (12)}~Radial velocity --302$\pm$2 \vkms~ (SDSS DR 9).~                    
{\bf (13)}~GG Aqr.~                               
{\bf (14)}~Faint object: Period and depth of minima uncertain. Drake et al. 
 (2014) give Period of 2.23825 days.~ Radial velocity +29$\pm$5 \vkms~(SDSS DR 9).~ 
{\bf (15)}~X-ray source (Flesch, 2010)             
{\bf (16)}~KM Boo.~                               
{\bf (17)}~Pojmanski (2002).~                             
{\bf (18)}~$(B-V)$ = 0.28; [Fe/H] =--1.56; Radial Velocity = --14$\pm$14 (Brown et al., 2008).~                             
{\bf (19)}~Radial velocites +76$\pm$6 \& +159$\pm$3 (SDSS DR 9).~               
{\bf (20)}~[C/Fe] = +3.4 (Christlieb et al., 2008).~                  
{\bf (21)}~Abbas et al. (2014).~                           
{\bf (22)}~Drake et al. (2014).~                           
{\bf (23)}~Drake et al. (2014) give type as EW and period of 0.339 days. See Sec. 4.6 for discussion.~                           
{\bf (24)}~Drake et al. (2014) give  same type and period of 0.583 days. See Sec. 4.6 for discussion.     \\                    
}
   \end{deluxetable*}

    We discover three new RRab of particular interest.  J100358.9-270001.4 has the 
    largest $(NUV-V)_{0}$ color and is the only object to lie in the color domain of 
    the disk RR Lyrae stars.  Yet at a heliocentric distance of roughly 5 kpc, it 
    must be about 2 kpc above the galactic plane.  Its proper motion of 8$\pm$2 mas 
    y$^{-1}$ (Girard et al., 2011) corresponds to a velocity of 200$\pm$50 km 
    s$^{-1}$.  This velocity and its relatively long pulsation period (0.6311 days) 
    suggest that J100358.9-270001.4 is not a disk star, but a halo star.  
    Spectroscopic radial velocity and metallicity measurements are needed to 
    establish its origin. The two RRab stars with the smallest $(NUV-V)_{0}$ colors 
    are J165153.4-223952.2 and J214038.1-233642.7.  J165153.4-223952.2 has the 
    largest E$(B-V)=0.419$ and therefore the most uncertain extinction correction of 
    the sample, while J214038.1-233642.7 has an unusually low $V$ amplitude for its 
    period and is possibly multiperiodic.

    The high amplitude \ds~stars in Table 1 are shown by green filled circles in 
    Fig. 6 and lie within the domain of the known \ds~stars which show a wide range 
    in $(NUV-V)_{0}$.  We note that the star SS Psc, which has a conflicting RRc 
    classification in the GCVS and a \ds~star classification by McNamara (1997), is 
    clearly in the \ds~star domain.

\section {The identification of RR Lyrae and \ds~Stars in the Gezari et al. (2013) Catalog.  }

    In Table 3 we present updated classifications and period measurements for 28 of 
    the 38 RR Lyraes stars identified by Gezari et al.\ (2013).  We find that 16 are 
    RRab, 4 are RRc, 6 are \ds~stars and 2 are eclipsing systems.  The main
    reason that we could not obtain acceptable light-curves for the remaining
    objects was because their visual amplitudes were comparable to the 
    observational errors of their magnitudes. A fuller discussion is given in
    Sec. 4.6. 

    Interestingly, Gezari et al. (2013) provide a ``light curve type" (LC) which is 
    either Stochastic (V) or Flaring (F) (col. 4, Table 3). We would expect that RR Lyrae 
    and \ds~stars to have ``Stochastic'' light curves, but two of the variables 
    that Gezari et al. type as RR Lyrae have ``Flaring" light curves in $NUV$.

    Another 17 variables in the Gezari catalog are identified simply as ``Stars.'' 
    We are able to make detailed classifications for 7 of these objects, listed at 
    the end of Table 3. We identify 1 RRab, 1 RRc and 5 eclipsing systems.  Five of 
    the ``Stars'' from the Gezari  catalog have ``Flaring'' light curves:  1 
    is an RRab, 2 are eclipsing systems and 2 are of unknown type.

    We provide ephemerides for the eleven newly identified variables from the 
    Gezari catalog in Table 2.

\begin{figure}
\epsscale{1.1}
\plotone{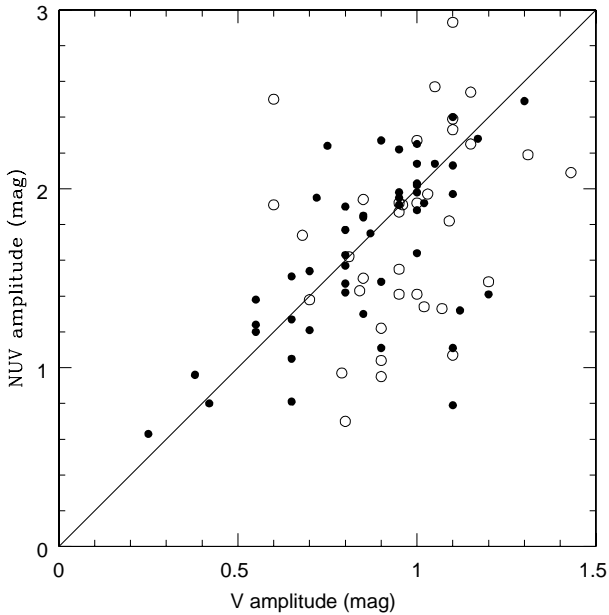}
\caption { The relation between the $NUV$ amplitude and the $V$ amplitude for
  84 type RRab variables. The line corresponds to the case where the $NUV$
  amplitude is twice the $V$ amplitude. The variables from Catalog 1 (open 
  circles) lie preferentially below the line while those from Catalog 2 (filled 
  circles) are more equally above and below the line.
                     }
\label{Fig7}
\end{figure}

\section{Discussion}

 We investigate the stellar properties of our sample of $UV$-selected variables.  
 We compare $NUV$ pulsation amplitudes with optical $V$ amplitudes, with the 
 distribution of pulsation periods, and with the distribution of non-variable stars. 
 We explore the completeness of the $UV$-selected sample, the problem of 
 distinguishing between contact binaries and pulsating stars and finally with
 a comparison of periods obtained in this paper with those independently
 obtained in a contemporous compilation by Drake et al. (2014). 

\subsection{The relation between the $NUV$ and $V$ amplitudes.}

 We begin by exploring the distribution of $NUV$ and $V$ pulsation amplitudes of the 
 $GALEX$ variables.  For context, consider the five RRab variables with 
 well-observed $NUV$ light curves from Wheatley et al. (2012). The mean of the ratio 
 of the $NUV$ amplitudes to the corresponding $V$ amplitudes (taken from the 
 literature) for these five RRab stars is 2.15$\pm$0.18. The [Fe/H] of these stars 
 range from $-1.5$ to $-1.7$ (mean $-1.64$), which means they are typical halo RR 
 Lyrae stars.

 In Figure 7 we plot the $NUV$ amplitude $vs.$ the $V$ amplitude for the 84 RRab 
 variables in Table 1.  The variables from Catalog 1 are plotted as open circles and 
 those from Catalog 2 by filled circles. The line represents the case where the 
 $NUV$ amplitude is twice that of the $V$ amplitude. In this figure, 11 of the RR 
 Lyrae from Catalog 1 lie above the line and 24 below; 26 of the variables from 
 Catalog 2 lie above the line and 23 below. The Catalog 1 variables lie 
 preferentially below the line (5\% level of significance) while those from Catalog 
 2 are more equally spaced.  Thus the $NUV$ amplitudes for the RRab derived from 
 Catalog 1 are on average lower than those from Catalog 2; the latter are generally 
 derived from more observations and approximate more closely to the true amplitudes. 
 The scatter in Fig. 7 is large so that the correlation between the $NUV$ and
 $V$ amplitudes is not strong. It is clear, however, that on average the
  RRab $NUV$ amplitudes are twice as large as their $V$ amplitudes.

\subsection{The $\log$ period-amplitude relation for $NUV$ and $V$ amplitudes.}

 Next we explore the period-amplitude relation of the variables. In Figure 8 we plot 
 the distribution of $\log$(period) versus $NUV$ amplitudes (left panel) and $V$ 
 amplitudes (right panel).  For reference, the curve shows the relation for 
 Oosterhoff Type I RRab variables observed in the globular cluster M3 (Cacciari et 
 al. 2008). The curve refers to monoperiodic variables; curves with secondary 
 periods such as RRd or those with Blazhko effect will have smaller amplitudes and 
 lie below the line in this plot. In the right hand plot, the variables with periods 
 shorter than 0.56 days ($\log$ P $\leq$ --0.25) lie mainly below the curve; those 
 with longer periods lie increasingly above the curve with increasing period. We 
 suspect that our $V$ amplitudes (derived by visual inspection from light-curves 
 largely derived from Catalina Survey data (Drake et al., 2009) or in a few cases  LINEAR Survey data (Sesar et al., 2013)) are systematically too 
 small and that the admixture of higher amplitude variables with periods greater 
 than 0.56 days comes from a population of Oosterhoff type II variables. The errors 
 in the amplitudes are too great for a clean separation of the two types. A similar 
 interpretation holds for the $NUV$ amplitudes in the left plot in Fig. 8 although 
 the scatter is larger.

 Two stars of interest are J013642.0-062743.1 and J171251.8+185452.4, which have 
 periods 0.7775660 and 0.8111936 days, respectively, and amplitudes less than 0.4 
 mag.  Schmidt (2002) has studied stars with periods between 0.60 and 1.0 days and 
 explored the various types of variables in this range. He finds 6 RRab with periods 
 greater than 0.75 days and only 4 with periods greater than 0.80 days. None of 
 these stars have amplitudes less than 0.50 mag. A new survey by Abbas et al. (2014) 
 finds $\sim$ 20 RRab with periods greater than 0.80 days. Most of these stars have 
 $V$ amplitudes between 0.4 and 0.6 magnitudes, and constitute about 0.03\% of their 
 total sample.  Here, we find two RRab with smaller amplitudes in this period range, 
 the aforementioned J013642.0-062743.1 and J171251.8+185452.4.  These two RRab 
 constitute $\sim$ 2\% of our sample, suggesting that the use of $GALEX$ magnitudes 
 (where observed amplitudes are larger) facilitates the discovery of such low 
 ampltiude stars.

\begin{figure}
\epsscale{1.2}
\plotone{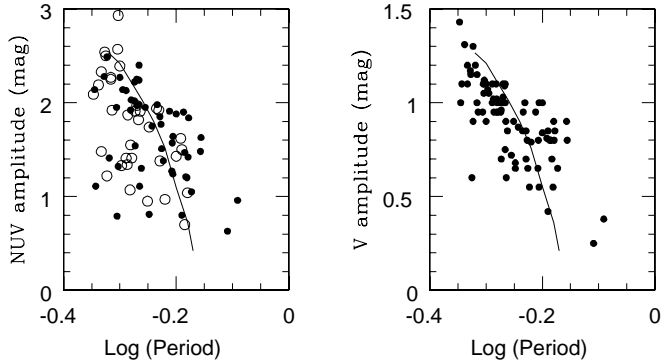}
\caption { Amplitude $vs.$ $\log$ period relation for the $NUV$ amplitude
   (left panel) and $V$ amplitude (right panel) for the 84 RRab variables from 
   Catalog 1 (open circles) and Catalog 2 (solid circles). The curve is the relation 
   found for the Oosteroff type 1 variables in the globular cluster M3 given
   by Cacciari et al. (2008); the amplitude of the curve has been doubled 
   for the $NUV$ amplitudes in the left panel.
                     }
\label{Fig8}
\end{figure}

\begin{figure}
\epsscale{1.0}
\plotone{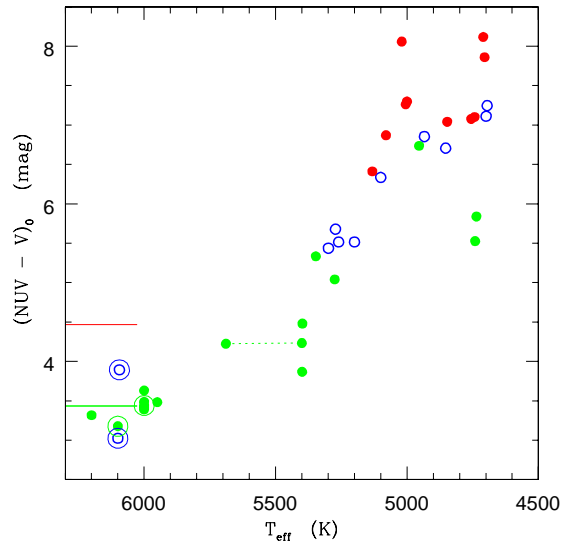}
\caption { $(NUV - V)_{0}$ color versus spectroscopic effective temperature 
  T$_{eff}$ for a selection of red horizontal branch (RHB) stars.  Colors indicate 
  metallicity:  [Fe/H] $>-0.80$ are shown by red filled circles, $-0.8 >$ [Fe/H] 
  $>-1.70$ are blue open circles, and [Fe/H] $<-1.70$ are green filled circles.  
  The two measurements for HD 119516 are joined by a dotted line.  Four stars 
  are optical variables and are indicated by encircled symbols (bottom left).
  The red and green horizontal lines show the mean values of$(NUV - V)_{0}$ for disk 
  and halo RRab variables (as shown in Fig. 6).  The data and notes on individual 
  stars are presented in Table A3 in the Appendix.
                     }
\label{Fig9}
\end{figure}

\subsection{The Red Horizontal Branch (RHB) stars. }

  In the plot of $(NUV - V)_{0}$ against period, the hottest RR Lyrae stars (type 
  RRc) are coterminous with the blue horizontal branch stars at the blue edge of the 
  instability gap. The stars near this blue edge are nearly all metal-poor; thick 
  disk blue horizontal branch stars are rare (Kinman et al. 2009).

  We now consider the non-variable horizontal branch stars on the cool side of the 
  instability gap: the red horizontal branch (RHB) stars. Fig. 9 plots $(NUV - 
  V)_{0}$ against effective temperature (T$_{eff}$) for these stars.  The sources 
  for the data in this plot are given in the Appendix (Sec. C). We see that the 
  hottest metal-poor RHB stars have $(NUV - V)_{0}$ colors similar to the halo RR 
  Lyrae stars. Indeed, we find that some of the hottest metal-poor RHB candidates 
  are actually RR Lyrae stars.

  For T$_{eff}$ less than about 5200 K, most of these RHB stars are metal-rich ([Fe/H] $>$ 
  -0.8) and have larger $(NUV - V)_{0}$ than the metal-poor RHB, similar to the 
  trend seen in the RRab stars.  We do not find any metal-rich RHB stars hotter 
  than this limit. As we discuss in Appendix C, this is likely an observational 
  selection effect; there are few constraints on hot metal-rich RHB stars at 
  fainter magnitudes ($V$ $>$ 10.5). We cannot rule out, however, that hot 
  metal-rich RHB stars may be scarce in the field even though they are found in the 
  metal-rich globular clusters like 47 Tuc.

  The change in $(NUV-V)_{0}$ color with effective temperature is much larger 
  compared with the change in optical color indices. Stars like BD 18$^{\circ}$ 2757 
  and BPS CS 22875-029 have similar metallicity but differ in T$_{eff}$ by 
  $\sim$1250 K. Their difference in $(B-V)_{0}$ and $(J-K)_{0}$ are 0.32 and 0.15 
  mag, respectively, while their difference in $(NUV - V)_{0}$ is over 2 mag. This 
  makes the $(NUV-V)_{0}$ color particularly sensitive for detecting the temperature 
  changes that come from pulsations and, in particular, for detecting the onset of 
  this instability as a function of other parameters (such as T$_{eff}$).

\begin{figure}
\epsscale{1.0}
\plotone{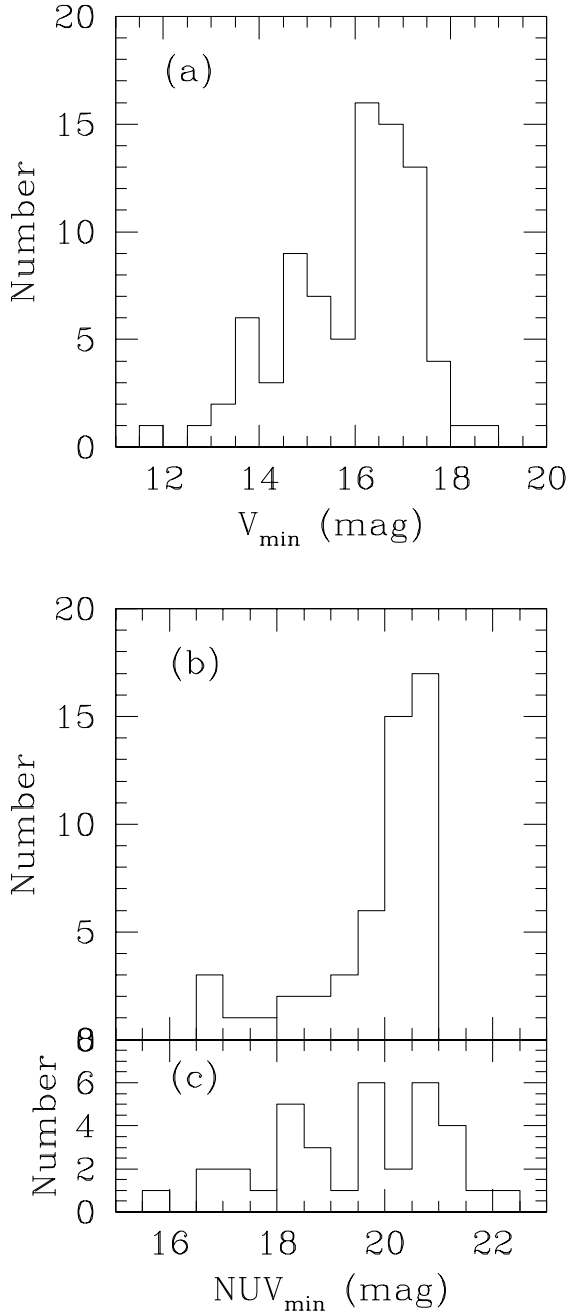}
\caption {  The distribution of the RRab in Table 3 versus their
  $V$ magnitude at minimum light (panel a).  The distribution versus $(NUV)$ 
  magnitude at minimum light for Catalog 2 (panel b) and Catalog 1 (panel c).
                     }
\label{Fig10}
\end{figure}

\subsection{The Completeness of the Samples.}

  We explore the magnitude limits of our samples in Figure 10.  Fig. 10(a) plots the 
  distribution of the minimum $V$ magnitudes for the 84 RRab variables in Table 1. 
  There is a steep fall-off in the distribution at $V$ $\sim$ 17.5.  Fig. 10(b) and 
  Fig. 10(c) plot the distribution of the faintest observed $NUV$ of the RRab in 
  Catalogs 2 and 1, respectively.  There is a steep fall-off at about $(NUV)$ = 21.0.  
  These two magnitude limits are consistent with the $3.4 \le (NUV - V) \le 3.8$ mag 
  RRab colors observed by Wheatley et al. (2012); most of the RRab in Table 1 are 
  also in this range.

  A $V$=17.5 magnitude corresponds to a distance limit of about 25 kpc for our 
  sample of RRab.  Most of the stars in our sample are at a sufficiently high 
  galactic latitude that the correction for interstellar extinction is small, 
  E$(B-V)$ $\leq$ 0.04 for which E$(NUV -V)$ $\leq$ 0.23.

  The mean $(NUV - V)$ for RRc is somewhat smaller than that of the RRab, and they 
  have lower amplitudes.  Consequently, RRc stars will have a fainter $V$ magnitude 
  for a given $NUV$, and their identification will necessarily be less complete than 
  for the RRab.  The $NUV$ light curve minima in the Gezari catalog are all fainter 
  than 20.0 and so, neglecting extinction, the RRc will have $V$ $>$ 17 and will 
  generally be less easy to detect with existing optical data without the use of new 
  techniques.

  The 17 RRc and 48 RRab corresponds to $\sim$0.40 variables per square degree for 
  Catalog 2. This is comparable to the 0.455 RR Lyrae per square degree found by 
  Abbas et al. (2014) in a survey of somewhat greater depth (28 kpc) and which they 
  find has a completeness of $\sim$50\%.  This suggests that $GALEX$ has detected 
  only half of the RRab to $V$=17.5, or else Catalog 2 is itself only about half 
  complete for these variables.
  
  Catalog 2 contains the variables whose $NUV$ amplitudes are greater or equal to 
  0.6 mag. The survey covered 169 fields; each field (radius 0.$^{\circ}$55) was 
  visited from between 10 and 187 times.  Table 4 shows that that variables were 
  only discovered in about 20\% of the fields when the number of visits was 20 or 
  less but this percentage rose to about 40\% when there were over 30 visits. The 
  field centered on AE Aqr is also in the well studied SDSS Stripe 82; three RRc and 
  2 RRab were found in this field but three RRc of comparable brightness that lie 
  within this field were missed. There were 31 visits to this field but the number 
  of actual detections of the RRc variables ranged from 19 for a 13th mag star to 5 
  for one at 17th mag. Thus the completeness will vary from field to field depending 
  on the number of visits. In the 19 fields that were visited more than 50 times, 
  eleven RR Lyrae variables were found and only two were missed.  It is therefore 
  only in these fields (that comprise only 11\% of the total) that the Catalog 2 
  survey is reasonably complete.

\subsection{Contact Binaries.}

 Contact binaries can be confused with RRc variables that have half their periods; 
 see discussions by Kinman \& Brown (2010) and Drake et al. (2014). Table 2 in the appendix (Sec. B) gives the $UV$ colors for a sample of bright
 contact binaries and these are plotted against period in Fig. 11. 

  Drake et al. (2014) use a variety of parameters to distinguish between RRc and 
  contact binaries and conclude that misidentification is most likely to occur among 
  the longer-period bluer contact binaries which are the brightest among such 
  variables; misidentification is therefore not a serious problem for faint surveys. 
  Misidentification apart, the cause of the spread in $(NUV - V)_{0}$ among contact 
  binaries needs further investigation. It is unlikely to be caused by a spread in 
  metallicity because the range of metallicity in these objects is small (Rucinski 
  et al. 2013). The presence of companions is a more likely cause since it is known 
  that multiplicity is very common among contact binaries (Pribulla \& Rucinski 
  2008). Smith et al. (2014) consider the effect of companion stars on $GALEX$ 
  colors in their discussion of stars in the Kepler field. $(NUV - V)_{0}$ may 
  therefore be a useful diagnostic in searching for such companion stars.

\begin{figure}
\epsscale{1.0}
\plotone{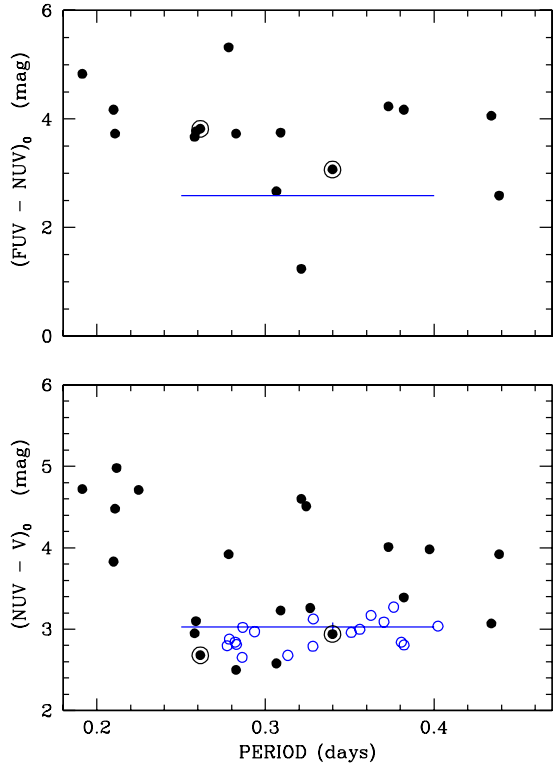}
\caption { $(FUV - NUV)_{0}$ (upper panel) and $(NUV - V)_{0}$ (lower panel) plotted 
  against the half-periods of a sample of bright contact binaries (black filled 
  circles).  Two of these objects (SW Ret and v1643 Sgr) that are shown encircled, 
  are actually RRc variables. The blue horizontal lines in both figures show the 
  location of RRc variables. Our program RRc variables are plotted as blue filled 
  circles using the full periods in the lower figure.
                     }
\label{Fig11}
\end{figure}

\begin{figure}
\epsscale{1.21}
\plotone{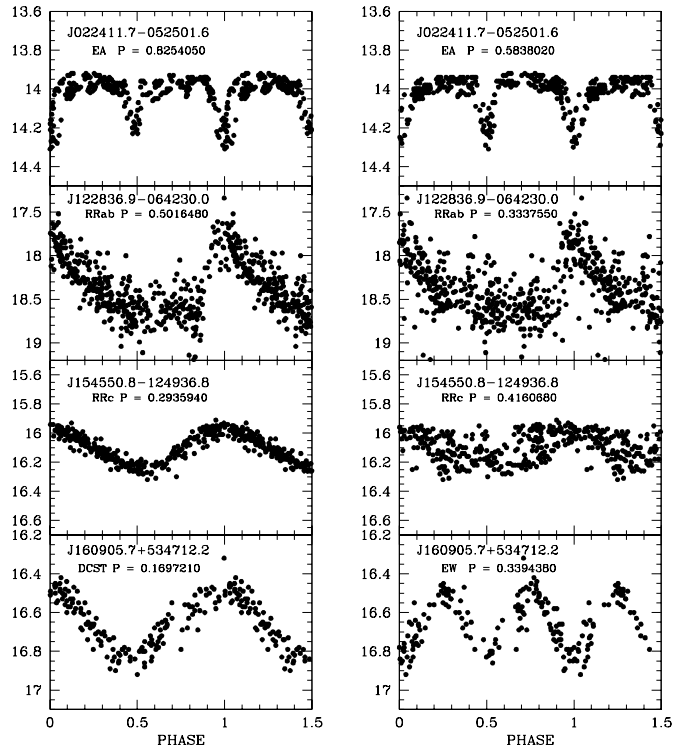}
\caption { Comparison of the light-curves of J022411.7-052501.6,
  J1222837.9-064230.0, J154550.8-124936.8 and 160905.7+534712.2 obtained 
  in this paper (on the left) and those given by Drake et al. (2014) 
  on the right.
                     }
\label{Fig12}
\end{figure}

\begin{deluxetable*}{ccc}
\tablewidth{0cm}
\tabletypesize{\footnotesize}
\tablecaption{ Discovery rate of RR Lyrae stars as a function of number of visits per field for Catalog 2.}
\tablehead{ 
\colhead{ Visits per Field     } &
\colhead{ No. of Fields with this } &
\colhead{ Number of Fields in which} \\
               &  range of visits&variables were found \\
 (1) & (2) &(3)    \\
}

\startdata
 10 to 20  &  86  &   18      \\   
 21 to 30  &  39  &   14      \\   
 31 to 50  &  25  &   11      \\    
  $>$ 51  &  19  &    8      \\    
\enddata

\end{deluxetable*}

\subsection{Uncertainties in determining Variable Type and Period.}

   Five of our new variables (Table 2) are also found in a new catalogue of periodic 
   variables from the Catalina Surveys (Drake et al. 2014, hereafter D14).  The 
   first star is J022121.9-033040.0, for which D14 find the same variable type and 
   period as we do.  Thus we do not discuss this star further.  For the other four 
   stars, we obtain repeat periodograms using the Lomb-Scargle (LS) and Plavchan 
   (PL) routines in the NASA Exoplanet Archive Periodogram Service. Fig. 12 presents 
   the light curves for these four variables.  For each variable, the left-hand 
   panel plots the data folded on our period, and the right-hand panel plots the 
   data folded on the D14 period.

   With the exception of J154550.8-124936.8, identifying the correct period and type 
   by visual inspection is difficult because of period alias uncertainties.  The 
   discrepant results described below illustrate the difficulty in identifying the 
   nature of low-amplitude variables.

     J022411.7-052501.6.  We identify this as an Algol variable with a period of 
     0.8254050 days.  D14 give the same type with a period of 0.583802 days.  The LS 
     routine gives neither of these periods.  The PL routine gives both periods; the 
     D14 period has more power in the PL routine and is likely correct.

     J122837.9-064230.0.  We identify this as an RRab with a period of
     0.501648 days.  D14 give the same type with a period of 0.333755 days, but
     call this period ``inexact.''  The LS and PL routines give both periods, 
     however our period has more power and is likely correct. 

     J154550.8-124936.8.  We identify this as an RRc with a period of 
     0.293594 days.  D14 give the same type with a period of 0.416068 days.
     Both the LS and PL routines give more power to our period than to the one at 
     0.416 days, and our period gives a better looking light curve (Fig. 12).

     J160905.7+534712.2.  We identify this as a \ds with a period of 0.1697210 days.  
     D14 identify it with an EW with a period of 0.339438 days.  The LP routine 
     gives our period.  The PL routine gives both periods with marginally more power 
     to our period.  However, the D14 identification is likely correct because the 
     $(g-r)=0.32$ color of the variable is probably too red for a \ds.

\section{Summary}

   We make as complete an identification as possible of unclassified variable stars 
   in existing $GALEX$ variable source catalogs.  We present newly identified \ds, 
   \dc~and \da~stars found in what we refer to as Catalog 1 (Welsh et al. 2005) and 
   Catalog 2 (Wheatley et al., 2008).  We also examine the sources identified either 
   as ``stars'' or RR Lyrae stars in the larger catalog of variable $GALEX$ sources 
   of Gezari et al. (2013).

   We identify 8 \ds~stars, 17 \dc~stars, 1 RRd star and 84 \da~ stars in Catalogs 1 
   and 2. These 110 pulsating variables account for 22\% of the sources in these 
   Catalogs. Sixteen of these variables were not previously known; their ephemerides 
   and light curves are given in Table 2 and Figures 1, 2, 3 and 4 respectively.

    In Table 3 we present new classifications and period measurements for 28 of the 38 
   variable $GALEX$ sources that Gezari et al (2013) identify as RR Lyrae stars:
   16 are \da~stars, 4 are \dc~stars, 6 are \ds~stars and 2 are eclipsing systems.  
   We also classify 7 of the 17 $GALEX$ sources that Gezari et al. identify as
   ``stars'': 1 is a \da~star, 1 is a \dc~star and 5 are eclipsing systems. The 
   classification of these objects depends on our ability to recognize their 
   variability type from optical data in the Catalina Surveys Data Release 2. 
   This becomes difficult for low-amplitude variables with $V>17.5$.  
   Thus classifications are less complete for the Gezari catalog because 
  the catalog contains lower amplitude variables ($NUV$ amplitude $\geq$ 0.2 mag
   compared with $\geq$ 0.6 mag for Catalog 2). The light-curves of the
   variables in Table 2 from the Gezari Catalog are given in Figures 1, 2, 3 
   and 4.

   We find that the $(NUV - V)_{0}$ color of low metallicity halo RRab stars is 
   smaller than that of the higher-metallicity $disk$ RRab stars. The $NUV$ 
   pulsation amplitude of RR Lyrae stars is roughly twice that of their $V$ 
   amplitude (as found earlier by Wheatley et al.  (2012)). This allows us to detect 
   low-amplitude variables such as J013642.0-062743.1 and J171251.8+185452.4 which 
   have $V$ amplitudes of 0.25 and 0.38 mag, respectively. A larger number might be 
   found if the Catalogs were more complete. At a rough estimate, the overall 
   completeness of Catalog 2 is similar to that of the catalog of Abbas et al. 
   (2014) which they estimate as 50\%. The completeness of Catalog 2 varies from 
   field to field depending on the number of observations made per field.

   $(NUV - V)_{0}$ is more sensitive than optical colors to changes in T$_{eff}$ (such 
   as those that occur in stellar pulsation).  Four stars that have been 
   described in the literature as RHB stars are actually RR
   Lyrae stars (e.g. BPS CS 22940-070 which is called non-variable by For et al., 
   (2010)). A plot of $(NUV - V)_{0}$ against T$_{eff}$ shows that the metal-poor
   RHB stars merge into the RR Lyraes at T$_{eff}$ $\sim$ 6000 K which agrees with
   other estimates of the red edge of the instability gap. Existing surveys such as
   Afsar et al. (2012) do not show any metal-rich (disk) RHB stars hotter than 
   T$_{eff}$ $\sim$ 5200 K whereas such stars are known in the metal-rich globular 
   cluster 47 Tuc. We suggest that hotter metal-rich RHB  stars could be found by 
   deeper surveys. 

   We find a large range in both $(NUV - V)_{0}$ and $(FUV - NUV)_{0}$ colors in 
   a sample of bright contact binaries that have twice the periods of RRc stars and
   which may therefore be confused with them.  These $UV$ colors do not 
   allow a clear distinction between the two types, but may be of value in detecting
   companions to the contact binaries.  

	~

\acknowledgments   

We acknowledge use of the International Variable Star Index (VSX) database 
operated at AAVSO, Cambridge, MA, USA. Also the LINEAR survey (available on
the SkyDOT website) and funded by NASA at MIT Lincoln Laboratory under Air Force
 Contract FA8721-05-0002. Also the Catalina Surveys (Data Release 2) that is
 supported by NASA under grant NNG05GF22G and NSF grants AST-0909182 and
  AST-1313422,
 We used the VizieR catalogue access tool, CDC, Strasbourg,
France. We acknowledge using data from the SDSS funded by the
Alfred P. Sloan Foundation, the Participating Institutions, the NSF and
the US D.O.E.. We also used the MAST 
 (Multimission Archive at the STSci which is operated
for NASA by AURA), the SIMBAD database (operated at the CDS, Strasbourg,
France), ADS (the NASA Astrophysics Data System), and the arXiv e-print
server. We also used the NASA Exoplanet Archive which is operated by the California
Institute of Technology under contract with NASA.

\clearpage

\appendix 

\section{High-Amplitude $\delta$ Scuti Stars (HADS).}

  The High-Amplitude $\delta$ Scuti stars are radially pulsating stars with
  periods in the range 0.04 to 0.30 days that are largely monoperiodic. 
  They show some overlap in period with the RRc variables but can usually
  be distinguished from them by having more asymmetric light curves. Table A1
  gives the $(NUV - V)_{0}$ for a sample of these stars drawn from the compilations
  of McNamara (1997) and Suveges et al. (2012). The HADS are well separated from 
  the RRc variables in the $(NUV - V)_{0}$ $vs.$ Period plot (Fig. 6); this is 
  clearly the case for SS Psc (which is marked in this figure). This is of 
  interest because Ferro et al. (2013) were unable to classify this star from
  their Stromgren photometry. The revised Hipparcos parallax of SS Psc is 5.26 $\pm$
  1.99 (van Leeuwen, 2007) gives an M$_{v}$ of between +3.5 and +5.2 which is in the
  expected range for HADS (Peterson \& Hog, 1998).

\begin{deluxetable*}{llcccc}
\tablewidth{0cm}
\tabletypesize{\footnotesize}
\setcounter{table}{1}
\tablecaption{$(NUV - V)_{0}$ for $\delta$ Scuti Stars.}
\tablehead{ 
\colhead{ R.A.  \tablenotemark{a}     } &
\colhead{ Dec.  \tablenotemark{b}     } &
\colhead{ Ref \tablenotemark{c}    }  &
\colhead{$(NUV-V)_{0}$ \tablenotemark{d}     } &
\colhead{ Period \tablenotemark{e}     } &
\colhead{ GCVS ID\tablenotemark{f}     } \\
 (deg.) & (deg) &   &  (mag.) & (days)   &            \\
 (1) & (2) &(3) & (4) & (5) & (6)   
}
\startdata
 144.2216 &+44.0668 & 1& 3.330 & 0.0860100 &  AE UMa  \\ 
 231.0294 &+36.8670 & 1& 3.118 & 0.1040900 &  YZ Boo  \\ 
 238.2908 &+06.0907 & 1& 3.193 & 0.1891500 &  CW Ser  \\ 
 013.8256 &+23.1637 & 1& 2.461 & 0.0786800 &  GP And  \\ 
 337.5121&--08.1075 & 1& 3.912 & 0.1683800 &  GV Aqr  \\ 
 247.8249 &+11.9979 & 1& 3.392 & 0.1486300 &  DY Her  \\ 
 247.5682 &+16.9185 & 1& 2.550 & 0.0946800 & v1116 Her\\ 
 357.1914&--08.1457 & 1& 3.767 & 0.1978230 &  BS Aqr  \\ 
 146.4452&--12.9940 & 1& 3.226 & 0.2233891 &  VX Hya  \\ 
 020.2182 &+21.7287 & 1& 3.898 & 0.2877928 &  SS Psc   \\ 
         &          &  &      &            &           \\
 270.6644 &+62.7186 & 2& 3.393 & 0.1969000 & v395 Dra  \\ 
 278.0269 &+40.5991 & 2& 2.790 & 0.1021400 & v593 Lyr  \\ 
 240.7758 &+26.2396 & 2& 4.357 & 0.1924460 &  BY CrB   \\ 
 237.4137&--76.4219 & 2& 2.182 & 0.0638630 & v360 Aps  \\ 
         &          &  &     &            &            \\
 343.0172& --0.8690& 3& 3.214& 0.0495655  &            \\
 320.2203&  +0.2755& 3& 2.067& 0.0500164  &            \\
 319.3871&  +0.8267& 3& 2.813& 0.0493681  &            \\
 318.9853& --0.8594& 3& 2.472& 0.0512525  &            \\
 044.3292& --0.6536& 3& 2.323& 0.0482597  &            \\ 
 037.4992&  +1.2482& 3& 2.380& 0.0509957  &            \\
 341.6031&  +1.1892& 3& 2.754& 0.0502367  &            \\ 
 338.5538&  +0.7835& 3& 2.794& 0.0537832  &            \\
 322.4356&  +1.0480& 3& 2.421& 0.0596027  &            \\ 
 320.6782&  +0.7816& 3& 2.386& 0.0538881  &            \\
 313.7465& --0.1246& 3& 2.024& 0.0515025  &            \\
 312.4206& --0.3539& 3& 2.749& 0.0931453  &            \\
 322.4695& --1.1719& 3& 2.989& 0.0805862  &            \\
 321.1715&  +0.0444& 3& 2.451& 0.0452487  &            \\
 043.1488&  +0.2955& 3& 2.480& 0.0550160  &            \\
 355.3285&  +0.7152& 3& 3.247& 0.1198865  &            \\
 323.7998&  +0.6225& 3& 2.782& 0.0872579  &            \\
 313.1301& --0.5333& 3& 2.554& 0.0530346  &            \\
 305.7112& --0.1257& 3& 2.468& 0.0521552  &            \\
 320.1692&  +0.9076& 3& 1.698& 0.0742401  &            \\
 316.8586&  +1.1715& 3& 3.410& 0.0678054  &            \\
 310.7717&  +0.0367& 3& 2.386& 0.0457611  &            \\
 310.3194&  +0.5634& 3& 2.739& 0.0663556  &            \\
 357.0749&  +0.9069& 3& 2.562& 0.0607696  &            \\
 314.2421&  +1.0685& 3& 2.722& 0.0702984  &            \\
 304.6618&  +1.0123& 3& 3.566& 0.0457270  &            \\
 346.1327&  +0.4127& 3& 2.390& 0.0565413  &            \\
 002.3078& --0.1846& 3& 2.426& 0.0519909  &            \\
 338.6537&  +0.3735& 3& 2.625& 0.0530301  &            \\
 320.8264&  +0.8867& 3& 2.186& 0.0706932  &            \\
 045.7345& --0.6974& 3& 2.804& 0.0581344  &            
\enddata


\tablenotetext{a}{Right Ascension in decimal degrees.}           
\tablenotetext{b}{Declination in decimal degrees.} 
\tablenotetext{c}{Reference for identification as $\delta$ Scuti star.} 
\tablenotetext{d}{Dereddened difference beteween mean $NUV$ and $V$ magnitudes.}
\tablenotetext{e}{Period in days.} 
\tablenotetext{f}{Identification in General Catalogue of Variable Stars.}  

\tablerefs{
{(1)}~McNamara (1997);~
{(2)}~Samus et al. (2012), (General Catalogue of Variable Stars);~
{(3)}~Suveges et al. (2012).}
 
\end{deluxetable*}

\clearpage

\section {Contact Binaries}

 Table A2 gives 
 the $(NUV - V)_{0}$ and $(FUV - NUV)_{0}$ for a sample of bright contact binaries 
 with half-periods between 0.19 and 0.44 days (primarily taken from Malkov et al., 
 2006).  We excluded variables at low galactic latitudes ($\mid$b$\mid <$ 
 14$^{\circ}$) in order to minimize errors in the corrections for extinction. The 
 extinction correction used to derive $(FUV - NUV)_{0}$ is less than that for $(NUV 
 - V)_{0}$, but $FUV$ is available for fewer stars. The data from Table A2 is 
 plotted in Fig. 11.

  The $(NUV - V)_{0}$ of the 17 RRc that we found in Catalogs 1 and 2 range 
  from 2.65 to 3.17 ($\sigma$ = 0.14 mag) while in a  sample of 50 RRc 
  from the Abbas et al. catalog, the range is from 2.36 to 3.52 ($\sigma$ = 0.26
  mag). The $mean$ $NUV$ in the former case were derived from multiple values 
  of $NUV$ whereas in the latter case only a single value was usually available;
  this may well explain the greater spread of $(NUV - V)_{0}$ found for the
  RRc in the Abbas et al. catalog. 
 
  Half the contact binaries in Table A2 have $(NUV - V)_{0}$ within the range that we 
  find for the RRc. Two of these stars with low $(NUV - V)_{0}$ are actually RR 
  Lyrae stars: SW Ret is an RRd (Szczgiel \& Fabrycky 2007) and v1643 Sgr is an RRc 
  (Dvorak 2004). The more recent catalog of Avvakumova et al. (2013) continues to 
  identify v1643 Sgr as a contact binary. The mean $(FUV - NUV)_{0}$ of 16 RRc 
  variables is +2.59$\pm$0.14. In Table A2, EK Aqr, AG Vir, v941 Her and the RRc 
  v1643 Sgr are within the range of $(FUV - NUV)_{0}$ occupied by RR Lyraes; AG Vir 
  is known to have an infra-red excess (Nichols et al. 2010).

\begin{deluxetable*}{cccccccc}
\tablewidth{0cm}
\tabletypesize{\footnotesize}
\tablecaption{ UV colors for bright Contact Binaries.}
\tablehead{ 
\colhead{ R.A.  \tablenotemark{a}     } &
\colhead{ Dec.  \tablenotemark{b}     } &
\colhead{ b     \tablenotemark{c}     } &
\colhead{$(NUV-V)_{0}$ \tablenotemark{d}     } &
\colhead{$(FUV-NUV)_{0}$ \tablenotemark{e}     } &
\colhead{ X-ray \tablenotemark{f}     } &
\colhead{ P$_{0.5}$ \tablenotemark{g}     } &
\colhead{ GCVS ID\tablenotemark{h}     } \\
 (deg.) & (deg) &(deg.)&      (mag.) & (mag) &    & (days)   &            \\
 (1) & (2) &(3) & (4) & (5)& (6) & (7)  & (8) 
}
\startdata
130.0071  &+18.9998 &+32.17 &4.72  & 4.83& X & 0.1914    &  TX Cnc  \\  
150.6998  &+01.0945 &+41.94 &3.83  & 4.17&...& 0.2099    &   Y Sex  \\
258.4909  &+16.3502 &+28.73 &4.48  & 3.73& X & 0.2108  &  AK Her  \\  
077.8104  &--08.5569&--26.13&4.98  & ... & X & 0.2117  &  ER Ori  \\  
141.6710  &--13.7518&+25.68 &4.71  & ... &...&0.2248  &  EZ Hya \\   
234.1083  &+15.5317 &+50.1  &2.95  & 3.67&...&0.2580  &  CC Ser \\   
299.2000  &+01.1000 &--14.0 &3.10  &3.77 &...&0.2588  &v0724 Aql \\
052.0838  &--64.9772&--44.8 &2.68  &3.82 &...&0.2614  &SW Ret$\dagger$    \\
126.1351  &--16.4032&+12.04 &3.92  & 5.32&...&0.2782  &  AV Pup  \\  
079.9754  &--35.9017&--33.2 &2.50  & 3.73&...&0.2826  & RZ Col   \\
354.8175  &--09.1534&--65.2 &2.58  &2.67 &...&0.3065  &  EK Aqr  \\  
160.1383  &+13.5669 &+56.59 &3.23  &3.75 &...& 0.3090  &  UZ Leo  \\
180.2646  &+13.0083 &+71.61 &4.60  &1.24 &...& 0.3213  &  AG Vir  \\  
143.0766  &--28.6278&+16.61 &4.51  &...  &...& 0.3242  &   S Ant  \\  
277.4200  &+31.0003 &+17.9  &3.26  &...  &...& 0.3266  &  IY Lyr  \\  
296.4417  &--40.9383&--27.3 &2.94  &3.07 &...& 0.3398  &v1643 Sgr$\dagger$ \\  
212.3643  &--15.5816&+43.3  &4.01  &4.23 &...& 0.3730  & CX Vir   \\
119.5459  &+72.7643 &+30.7  &3.39  &4.17 &...& 0.3821  & CD Cam   \\
311.6156  &--71.9495&--34.5 &3.98  &...  &...& 0.3974  & MW Pav   \\ 
274.2226  &+27.6628 &+19.3  &3.07  &4.06 &...& 0.4340  & MS Her   \\
252.3800  &+47.1080 &+39.9  &3.92  &2.59 &...& 0.4386  &v921 Her
\enddata

\tablecomments{~~$\dagger$~~Both SW Ret and v1643~Sgr are RR Lyrae stars.}

\tablenotetext{a}{Right Ascension in decimal degrees.}          
\tablenotetext{b}{Declination in decimal degrees.}
\tablenotetext{c}{Galactic Latitude.}
\tablenotetext{d}{Dereddened difference beteween mean $NUV$ and $V$ magnitudes.}
\tablenotetext{e}{Dereddened difference beteween mean $FUV$ and $NUV$ magnitudes.}
\tablenotetext{f}{Given as X-ray source by Flesch (2010).}
\tablenotetext{g}{Half of period in days.}
\tablenotetext{h}{Identification in General Catalogue of Variable Stars.}  

 \end{deluxetable*}

\clearpage

\section { Red Horizontal Branch (RHB) Stars. }

 Fig. 9 (Sec. 4.3) shows the trend of $(NUV - V)_{0}$ with effective temperature 
 (T$_{eff}$) for RHB stars in three metallicity ([Fe/H]) ranges.  The data used in 
 this plot are given in Table A3.  We take all stars with the type RHB in Table 2 of 
 Behr (2003), and we also take NSV 4256 (HD 82590), which Behr lists as RR?, because 
 its T$_{eff}$ (6094 K) is near the red edge of the instability gap.  The 
 classification of NSV 4256 is uncertain but it is likely to be one of the lowest 
 amplitude RR Lyrae stars known; both Corben et al. (1972) and Alfonzo-Garzon et al. 
 (2012) give NSV 4256 a $V$ amplitude of 0.06 mag. Further data were taken from: 
 Preston et al. (2006), For \& Sneden (2010) and Afsar et al. (2012).  We expect 
 that errors in T$_{eff}$ are about $\pm$150 K. To illustrate this uncertainty, we 
 provide data for HD 119516 from two different sources in Table A3, and plot the 
 points separately (joined by a dotted line) in Fig.~9.

 Behr (2003) place the red edge of the instability strip at T$_{eff}$=6000 K (as 
 plotted in Fig. 9). Preston et al. (2006) and For \& Sneden (2010) give 6310 $\pm$ 
 145 K and 5900 K, respectively, for this quantity. These temperatures are 
 appropriate for the lowest metallicity stars (plotted in green in Fig 5.).  
 Bearing in mind the uncertainties in T$_{eff}$, these estimates for the red edge of 
 the instability strip are in reasonable agreement with our data.  We find that 
 $(NUV - V)_{0}$ increases with increasing metallicity at a given T$_{eff}$.

 There is, however, a large gap in T$_{eff}$ between the hottest metal-rich RHB 
 stars that we have found in the literature and the T$_{eff}$ of the metal-rich disk 
 RR Lyrae stars.  We estimate the latter to have T$_{eff}$ $\geq$ 6500 K based on 
 the $(V-K)$ of a small sample of these stars. The blue tip of red horizonal branch of the metal-rich 
 globular cluster 47 Tuc (Salaris et al. 2007) has $J - K$ $\sim$ 0.40 to 0.45,
 whereas the hottest of our metal-rich RHB stars has $J - K$ = 0.53. We therefore 
 suspect that the lack of hotter stars among our sample of metal-rich RHB stars is 
 an observational selection effect. The faintest 15 of the 76 stars studied by 
 Afsar et al. (2012) have $V$ between 9.0 and 10.5, which for M$_{v}$ =
 +0.8 corresponds to distances between 436 and 870 pc. Thus, as they point out, 
 their sample strongly favors thin disk stars. A search for thick disk RHB stars must be 
 made at significantly fainter magnitudes.

\begin{deluxetable*}{lccccc}
\tablewidth{0cm}
\tabletypesize{\footnotesize}
\tablecaption{$(NUV - V)_{0}$, [Fe/H] and T$_{eff}$ for a sample of RHB stars.}
\tablehead{ 
\colhead{ Identification    } &
\colhead{$(NUV-V)_{0}$ \tablenotemark{a}     } &
\colhead{ [Fe/H]    } &
\colhead{ T$_{eff}$ } &
\colhead{ Ref. } &
\colhead{ Notes   } \\
       &  mag. &     & $\deg$ K &     &                 \\
 (1) & (2) &(3) & (4) & (5)& (6)  
}
\startdata
 HIP 62325            & 8.117 & --0.06 & 4710 &  (6)  &        \\ 
 BD 25$^{\circ}$2459  & 7.101 & --0.28 & 4743 &  (1)  &        \\  
 HIP 115839           & 6.285 & --0.40 & 5000 &  (6)  &        \\  
 BD 34$^{\circ}$2371  & 7.262 & --0.41 & 5005 &  (1)  &        \\  
 HIP 3031             & 8.058 & --0.53 & 5020 &  (6)  &        \\  
 BD 29$^{\circ}$2294  & 6.411 & --0.55 & 5132 &  (1)  &        \\  
 BD 29$^{\circ}$2231  & 7.076 & --0.65 & 4756 &  (1)  &        \\  
 HIP 45412            & 6.870 & --0.68 & 5080 &  (6)  &        \\   
 BD 25$^{\circ}$2436  & 7.041 & --0.76 & 4847 &  (1)  &        \\  
 BD 36$^{\circ}$2303  & 7.860 & --0.77 & 4705 &  (1)  &        \\  
 HIP 5104             & 6.335 & --0.86 & 5100 &  (6)  &        \\
 HD~110930            & 6.854 & --0.94 & 4934 &  (1)  &        \\  
                      &       &        &      &       &        \\
 HIP 4960             & 5.516 & --1.02 & 5260 &  (6)  &        \\   
 HD~112030            & 7.111 & --1.12 & 4699 &  (1)  &        \\  
 HD~202573            & 6.706 & --1.21 & 4853 &  (1)  &        \\  
 BD 27$^{\circ}$2057  & 7.246 & --1.25 & 4695 &  (1)  &        \\  
 HD~6229              & 5.516 & --1.35 & 5200 &  (1)  &        \\  
 HD~97                & 5.678 & --1.42 & 5272 &  (1)  &        \\  
 BPS CS 22940-070     & 3.024 & --1.47 & 6100 &  (2)  &{\bf (1)} \\  
 HD~82590             & 3.893 & --1.50 & 6094 &  (1)  &{\bf (2)} \\  
 HD~105546            & 5.437 & --1.67 & 5299 &  (1)  &        \\  
                      &       &        &      &       &        \\
 HD~63791             & 6.735 & --1.72 & 4954 &  (1)  &        \\  
 BD 18$^{\circ}$2890  & 5.334 & --1.78 & 5347 &  (1)  &        \\  
 HD~119516            & 4.224 & --1.92 & 5689 & (1)(4)&        \\  
 BD 10$^{\circ}$2495  & 5.040 & --2.07 & 5275 &  (1)  &        \\  
 BD 17$^{\circ}$3248  & 4.480 & --2.08 & 5398 &  (1)  &        \\  
 BPS CS 22886-043     & 3.631 & --2.15 & 6000 &  (2)  &        \\ 
 HD~119516            & 4.233 & --2.16 & 5400 &  (3)  &        \\  
 BPS CS 22888-047     & 3.487 & --2.35 & 6000 &  (2)  &{\bf (3)} \\ 
 BD 18$^{\circ}$2757  & 5.527 & --2.43 & 4741 &  (1)  &        \\  
 BPS CS 22882-001     & 3.483 & --2.54 & 5950 &  (3)  &        \\ 
 BPS CS 22875-029     & 3.391 & --2.66 & 6000 &  (3)  &        \\ 
 BPS CS 22191-029     & 3.417 & --2.72 & 6000 &  (3)  &        \\ 
 BPS CS 22881.039     & 3.178 & --2.73 & 6100 &  (3)  &{\bf (4)} \\ 
 BPS CS 22186-005     & 3.318 & --2.77 & 6200 &  (3)  &{\bf (5)} \\ 
 BPS CS 30317-056     & 3.441 & --2.85 & 6000 &  (5)  &{\bf (6)}
\enddata


\tablenotetext{a}{Dereddened difference beteween mean $NUV$ and $V$ magnitudes.};

\tablerefs{
{(1)}~Behr (2003);~
{(2)}~Preston et al. (2006);~
{(3)}~For \& Sneden  (2010). 
{(4)}~Ishigaki et al. (2013);~   
{(5)}~Hansen et al. (2011). 
{(6)}~Afsar et al. (2012). }
 
\tablecomments{
{\bf (1)}~ RRab variable; Period = 0.5103 days; $V_{amp}$$\sim$ 0.9 mag. (this paper) 
{\bf (2)}~ NSV 4526; $V_{amp}$ $\sim$0.06 mag. (Corben et al. 1972; Alfonzo-Garzon et al.
, 2012); type uncertain. 
{\bf (3)}~[C/Fe] = 1.4 (Christlieb et al., 2008).
{\bf (4)}~RRab variable, Period = 0.66876 days; $V_{amp}$$\sim$0.8 mag. (Hansen et al.,
      2011);~[C/Fe] = 2.0 (Christlieb et al., 2008) 
{\bf (5)}~[C/Fe] = +1.0 (Christlieb et al., 2008) 
{\bf 6}~RRab variable, Period = 0.74851 days; $V_{amp}$$\sim$0.35 mag. (Hansen et al., 2011) 
}
\end{deluxetable*}

\end{document}